\documentclass{amsart}
\usepackage{multirow}
\usepackage{rotating}
\usepackage{amsthm}
\usepackage{amsfonts}
\usepackage[UKenglish]{babel}
\usepackage{color}
\usepackage{graphicx}
\usepackage{soul}
\usepackage{marginnote}
\usepackage{subfig}
\usepackage{stfloats}
\usepackage{fullpage}
\usepackage{morefloats}
\usepackage{setspace}
\usepackage{cite}
\newtheorem{theoremm}{Theorem}

\newtheorem{assumptionB}{Assumption}
\theoremstyle{lemma}

\theoremstyle{plain}

\theoremstyle{definition}

\theoremstyle{remark}

\begin{document}

\begin{twocolumn}
\title{Predicting Knot and Catenane Type of Products of Site-specific Recombination on Twist Knot Substrates}
\author{Karin Valencia$^a$ and Dorothy Buck$^b$}
\date{\today}
\maketitle
\begin{center}
\small{(A) (corresponding author) e-mail: karin.valencia06@imperial.ac.uk\\Imperial College London, South Kensington Campus, Department of Mathematics, Office: 640 London SW7 2AZ, England\\ Telephone number: +44 (0)20 758 58625\\(B) d.buck@imperial.ac.uk}\\Imperial College London, South Kensington Campus, Department of Mathematics, Office: 623, London SW7 2AZ, England
\end{center}

\small
\section*{Abstract}
{\footnotesize{Site-specific recombination on supercoiled circular DNA molecules can yield a variety of knots and catenanes. Twist knots are some of the most common conformations of these products and they can act as substrates for further rounds of site-specific recombination.  They are also one of the
simplest families of knots and catenanes. Yet, our systematic understanding of their implication in DNA and important cellular processes like site-specific recombination is very limited. Here we present a topological model of site-specific recombination characterising all possible products of this reaction on twist knot substrates, extending previous work of Buck and Flapan.  We illustrate how to use our model to examine previously uncharacterised experimental data. We also show how our model can help determine the sequence of products in multiple rounds of processive recombination and distinguish between products of processive and distributive recombination. 

This model studies generic site- specific recombination on arbitrary twist knot substrates, a subject for which there is limited global understanding. We also provide a systematic method of applying our model to a variety of different recombination systems.}}


\section{Introduction}\label{BioBackgroundandMotivation}

\subsection{Site-specific recombination}\label{SSR}


Site-specific recombination is a cellular process that involves reciprocal exchange between defined DNA sites. Prototypes of site-specific recombination include the integration of bacteriophage $\lambda$ into the \textit{Escherichia coli chromosome} and the DNA inversions responsible for flagellar phase variation in Salmonella \cite{bioreview}.  Apart from their fundamental functions in the cell, site-specific recombinases give scientists an elegant, precise and efficient way to insert, delete, and invert DNA segments. This means that they are rapidly becoming of pharmaceutical and agricultural interest and are being used in the development of biotechnological tools \cite{mouse, tools, MoreBiotoolsExamples1, MoreBiotoolsExamples2}.

\

Minimally, site-specific recombination requires one or two duplex  DNA molecules (linear, relaxed and plectonemically supercoiled covalently closed-circular  DNA are all good substrates for many site-specific recombination reactions. However, plectonemically supercoiled covalently closed circular DNA molecules are the most prevalent in topological enzymology studies, so we focus on these substrates here)  containing two short (30-50 base pairs (bp)) sequence specific DNA segments, the \textit{crossover sites} and specialized proteins,  \textit{site-specific recombinases}, responsible for recognizing the sites and  breaking and rejoining the DNA with conservation of the phosphodiester bond energy (Figure \ref{SSRsummary}). The sites are usually nonpalindromic, so each can be assigned an orientation and if the sites are on a single DNA molecule, they can either be in \textit{direct} orientation (head-to-tail) or in \textit{inverted} orientation (head-to-head). Depending on the initial arrangement of the parental recombination sites and recombinase used, site-specific recombination has one of three possible outcomes: integration, excision or inversion.  Larger site-specific recombination systems may also require additional proteins (e.g., accessory proteins) and sites (e.g., accessory sequences).

\

 {The reaction starts when a recombinase dimer binds at each of the two recombination sites (from now on, sites). The sites are then brought together to form the \textit{synaptic complex} with the crossover sites juxtaposed, possibly trapping a fixed number of (interdomainal) supercoils.} The sites are cleaved, exchanged and resealed. Finally, the proteins dissociate releasing the product molecule, completing the reaction. (Figure 1.) 

\

We refer to the region of space containing the two juxtaposed sites and the recombinase molecules as the \textit{recombinase complex}. 
 The synaptic complex is called a \textit{productive synapse} if the recombinase complex meets the substrate in precisely the two crossover sites. In this model (as opposed to the tangle model), we assume that any enhancer sequences and/or accessory proteins are sequestered from the recombinase complex and that the recombinase complex meets the substrates at precisely the two crossover sites. That is, we assume that the synaptic complex is a productive synapse (Figure \ref{prodsynp}). During the intermediate step, once the crossover sites have been cleaved, multiple rounds of strand exchange can occur before resealing the DNA, this is called \textit{processive recombination}. The entire process of recombination (including releasing and rebinding) can also occur multiple times, either at the same site or at different sites, this process is called \textit{distributive recombination}. In this work we use the term \textit{substrate} to refer specifically to the DNA prior to the first cleavage. Processive recombination is treated as one extended process, given an initial substrate with several intermediate exiting points for the reaction. 

\

Site-specific recombinases can be broadly divided into two subfamilies: serine recombinases and tyrosine recombinases,  based on sequence homology, catalytic residues and their mechanisms of cutting and rejoining the DNA. Only serine recombinases can mediate processive recombination. See \cite{bioreview} (and references therein) for a detailed exposition of site-specific recombination.

\subsection{DNA knots and catenanes}\label{DNAKnotsCatenanes}

A variety of DNA knots and catenanes have been observed since their discovery in the 1960s \cite{circularDNA1, circularDNA2 , circularDNA3, knottedEcoliDNA}. (Experimentally, two techniques have been widely used to resolve DNA knots and catenanes, electron microscopy and electrophoretic migration, \cite{14,15,16}.) 
However, they arise more commonly as products of topological enzymology experiments on artificially constructed small (3-5 kb) DNA plamids. Knots and catenanes are generated and then used as experimental tools to investigate the mechanisms of enzymes acting on DNA. They are used as substrates for these reactions and careful analysis of the topology of the product DNA molecules allows inferences to be drawn about the detailed mechanism of the reaction \cite{vazquez-gin, Sumners-et-al,vazquez-et-al, k1, k2, Darcy,d1, d2, d3, d4, d5, Kanaar1990, 8, 5, 6, 7, 12, 13, 14, ernst1, ernst2, 3tangle,v1,topoice,Buckintegrases, 10, Bint, 31,BuckMarcotte,BuckMauricio, 31, Darcy}. 


\subsection{DNA twist knots as substrates for site-specific recombination}\label{twistknotsubstrate}
A \textit{twist knot $C(2,v)$} is a knot that admits a projection as illustrated in Figure \ref{substrate}. Mathematically, twist knots are the simplest family of knots (after the torus knots and catenanes $T(2,m)$, which admit a projection as in Figure \ref{TorusKnott}) and appear more prevalently for small \textit{minimal crossing number} (See Section 2.1 for a definition).

\

Twist knots are ubiquitous DNA knot molecules \textit{in vivo} and \textit{in vitro}. Most DNA inside prokaryotic cells is plectonemically supercoiled and in the lab most experiments done with site-specific recombinases use small plectonemically supercoiled circular DNA molecules. This supercoiling promotes strand collision and DNA entanglement. A simple crossing change in such a molecule can result in knotting of the DNA into twist knots (Figure  \ref{supercoiledtotwist}). 

\

 Together with torus knots and catenanes, twist knots are the most common products of site-specific recombination both \textit{in vivo} \cite{6} and \textit{in vitro} \cite{4,5,6,7,8,10}, mediated by serine recombinases and tyrosine recombinases on unknotted, unlinked and torus knot and torus catenane substrates (see Table 1 in \cite{BFbio} and references therein).  For example, recombination mediated by $\lambda$ Int on the torus catenane $T(2, -2)$ with \textit{attP} and \textit{attB} bacteriophage lambda attachment sites, one on each of the components of the catenane substrate, yields the twist knot products $C(-2, 3),$ $C(-2, 5),$ $C(-2, 7)$ and $C(-2, 9)$ (Table 7 in \cite{10}, see also the top image of Figure \ref{Page3Exampless}).
Site-specific recombination mediated by Hin recombinase on an unknot substrate with inverted sites, has the following sequence of processive recombination:
$C(-2, -1)$ (substrate) $\rightarrow$ $C(-2,-1)$ $\rightarrow$ $C(-2,1)$ $\rightarrow$ $C(-2, 2)$ $\rightarrow $ $C(-2, 3)$  \cite{6} (see also the middle image of Figure \ref{Page3Exampless}). The twist knots $C(-2,1)$,$C(-2, 2)$ and $C(-2,3)$ are products of the second, third and fourth rounds of processive recombination on an unknotted substrate.

\

 Experimental conditions do not always preclude distributive rounds of recombination, and both can occur\footnote{Distributive recombination can be minimized  for example by stereostructural impediments or diluted  protein concentration (see e.g., \cite{5, johnson-bruist}).}. In multiple rounds of processive and distributive recombination on unknot, unlink and torus knot and catenane substrates, twist knots can become substrates of new recombination reactions. 
For example, in experiments of Kanaar et al \cite{Kanaar1990}, site-specific recombination on an unknot substrate with inverted sites, mediated by Gin, yields the following products of processive recombination: $C(-2,-1)$ (substrate) $\rightarrow$ $C(-2,-1)$ $\rightarrow$ $C(-2,1)$ $\rightarrow$ $C(-2, 2)$ $\rightarrow $ $C(-2, 3)$  \cite{6} (see also the middle image of Figure \ref{Page3Exampless}). The twist knots $C(-2,1)$,$C(-2, 2)$ and $C(-2,3)$ are products of the second, third and fourth rounds of processive recombination.  Also, the composite knot on six crossings, the granny knot $C(-2,1)\sharp C(-2,1)$ was analysed to be a product of distributive recombination on two trefoils, each a product from the first round of recombination.

\

Thus, a better understanding of DNA twist knots and their role in site-specific recombination reactions may contribute to the understanding of the mechanisms of this  cellular process.

\subsection{Our Model}
 Given the variety of DNA knots and catenanes that arise from site-specific recombination, it is clear that stratification of these products is necessary. Topological techniques such as those presented here, can aid experimentalists in characterizing DNA knot and catenane products. Our model predicts the exact topology and chirality of possible products, thus restricting the knot or catenane type of the products observed.

\

Despite the ubiquity and biological importance of of these knots, previous systematic study of twist knots involved in DNA-protein interactions has been limited and there has yet to be a systematic model incorporating
these as substrates for a generic site-specific recombinase. Earlier predictions of knots arising from site-specific recombination did not consider twist knot substrates \cite{Sumners-et-al, BFmaths, BFbio}. Here we rectify this by presenting a
model, extending the work of \cite{BFmaths}, classifying all possible knots and links that can arise from site-specific
recombination on a twist knot substrates.

\

 Our model is built on three biological assumptions,  stated in Section \ref{assumptions}.   From these, we construct a model that predicts all possible knots and catenanes that can arise as products of a single round of recombination, multiple rounds of (processive) recombination, or distributive recombination on a plectonemically supercoiled twist knot substrate $C(2, v)$. We predict that products arising from site-specific recombination on a twist knot substrate $C(2, v)$ must be members of one family of products, illustrated in Figure \ref{fams}$(a)$. Our model is independent of site orientation. We make no assumption on the size (number of basepairs) of the molecule(s). In \cite{KDmaths} we provide detailed topological proofs for the model presented here.

\

Note that, although the three assumptions are not for a generic recombinase, but rather, for general site-specific recombination, our model can restrict the topology of products in a specific system.  If for example, the site-orientation is taken into account, the model further restrict the possible products of such a reaction. We illustrate this in Section \ref{applications} with many examples.

\subsection{Structure of this paper}

This article is organized as follows.  In Section \ref{backgroundterminology} we explain  mathematical terminology and  notation. In Section \ref{assumptions}, we state the three assumptions of our model. In Section \ref{results}, we explain how, given a twist knot $C(2,v)$, all possible knotted or catenated products fall into one characterised family with two important subfamilies. We also consider the (common) case of products that have minimal crossing number (MCN) one more than the substrate, and show that the product knot or catenane type is even more tightly prescribed.  Finally, in Section \ref{applications}, we discuss how the model can help predict all possible products of (non-distributive) recombination mediated by a serine recombinase and a tyrosine recombinase, determine the order of products of processive recombination, distinguish products of distributive recombination, and narrow the possible knot or catenane type for previously uncharacterised experimental data.


\section{Mathematics Terminology and Notation}\label{backgroundterminology}

 In this section we define a few mathematical terms and introduce notation. Figures   \ref{prodsynp} and \ref{defs}   present diagrams for each one of the  terms defined. (We note that all line segments in these images represent the central axis of the double helix of a duplex DNA molecule.) 

\

Throughout this article, we  adopt the convention for crossings illustrated in Figure \ref{crossingconvention}. 

\subsection{Mathematical terminology}\label{mathematical_terminology}

 A \textit{catenane} $L$ is a collection of separate rings that may or may not be knotted, called \textit{components}. (Figure \ref{linkknot}.)
 A \textit{knot} $K$ is considered to be a catenane of one component (Figure \ref{linkknot}.)
Roughly, two knots or catenanes $K, J$ are \textit{equivalent} if there is a continuous deformation from $K$ to $J$ (without cutting them). A \textit{torus knot or catenane}, $T(2, m)$ is a knot or catenane formed by closing a row of $m$ plectonemic twists (such a closing is achieved by identifying the top and bottom endpoints of the arcs representing twists, without introducing or removing further crossings, see Figure \ref{TorusKnott}).  
 A \textit{twist knot $C(2,v)$} is a knot that admits a projection with two non-adjacent rows of crossings: a row of $v\neq0,1$ vertical crossings and a \textit{hook}, of the form shown in Figure \ref{substrate}. Note that $C(-2,v-1)=C(2,v)$. The equivalence of these two forms can be seen via a continuous deformation by flipping the hook. Note that twist knots can be generalized  to \textit{clasp knots} (Figure \ref{claspknot}). A clasp knot $C(r,v)$ is a knot that has two non-adjacent rows of crossings, one with $r\neq0,\pm1$ crossings and the other with $v\neq0$ crossings. A clasp knot $C(r, v)$ with $r = \pm2$ is a twist knot.

 \
 
In this work we may use the standard Rolfsen notation and the $T(2,m)$, $C(2,r)$ notation interchangeably. We give examples of both notations for knots and catenanes with the smallest minimal crossing number (the standard Rolfsen notation is on the left): $0_1=C(2,1)$, $(+)2^2_1=T(2,2)=C(2,0)$, $(-)3_1=C(-2,1)$, $(+)3_1=C(2,-1)$,  $4_1=C(2,-2)$, $(-)4^2_1=T(2,-4)$, $(+)5_1=T(2,5)$, $5_2=C(2,-3)$, $6_1=C(2,-4)$. For more examples, see \cite{Rolfsen}.  Note that $T(2, m)$ is a catenane if $m$ is even and is a knot if $m$ is odd (Figure \ref{TorusKnott}).

\

Given two knots or catenanes $K_1$ and $K_2$, their \textit{composite knot or catenane}, written $K_1\sharp K_2$, is obtained by removing an unknotted arc from each and gluing the resulting two endpoints of $K_1$ to the two endpoints of $K_2$ without introducing (or removing) any additional knotting (Figure \ref{compositeknot}). 
A \textit{prime} knot is one that can only be decomposed into two sub-knots $K_1 \sharp K_2$ if one is trivial (i.e. equivalent to the unknot). 
The \textit{minimal crossing number }of a knot or catenane $K$, denoted MCN$(K)$, is the fewest number of crossings with which it can be drawn. For example, MCN (unknot) $=0$ and MCN$(C(2,-2))=4$. Similarly, MCN$(C(2,v))$$=$$|v|+2$ if $v<0$ or MCN$(C(2,v))$ $=$$v+1$ if $v>0$.  See \cite{Rolfsen, Cromwell, Kawauchi} for a mathematical study of knots and catenanes.

\

 Let $J$  denote the substrate $C(2,v)$.
 Recall that in this model  we assume that the synaptic complex is a productive synapse (defined in page 2, see also Figure \ref{prodsynp}), so let $B$ denote the smallest region containing the four bound recombinase molecules and the two crossover sites, that is, the recombinase complex. ($B$ is a topological ball i.e., it can be continuously deformed to a round ball, see Figure \ref{SSRsummary}.)
 The \textit{recombinase-DNA complex}  $B\cup J$ is the recombinase complex $B$ along with the rest of the substrate molecule $J$.

\subsection{Notation for product families}\label{familiesexplanation}

In the Section \ref{results} we show all knots and catenanes arising from site-specific recombination on a twist knot substrate must fall into family $F(p,q,r,s,t,u)$, containing subfamilies $G_1(k)$ and $G_2(k)$, illustrated in Figures \ref{famsF}, \ref{famsG1} and \ref{famsG2} respectively.

\

 In the family $F(p,q,r,s,t,u)$ of knots and catenanes, the variables $p,q,r,s,t,u$ describe the number of crossings between two DNA duplexes in that particular row of crossings. In this family, the variables $p,$$ q,$$ r,$$ s,$$ t,$$ u$ can be positive, negative or zero. $t,r$ and $p$ can take horizontal and vertical zero crossings and only horizontal non-zero crossings. $u,s$ and $q$ can only take (both zero and non-zero) vertical crossings. 
%
In the subfamilies $G_1(k)$ and $G_2(k)$ of knots and catenanes, the variable $k$ describes the number of crossings between the two DNA duplexes. Depending on the value of $k$, we obtain either a knot or a catenane: if $k$ is odd, the members of these families are knots and if $k$ is even, then the members of these families are two-component catenanes.  (See Algorithm 3, point $(3)$ in Discussions and Applications.)
Note that there are knots and catenanes that have projections in both $F(p, q, r, s, t, u)$ and one of $G_1(k)$ or $G_2(k)$. For example the trefoil knot has a projection as a member of $F(p,q,r,s,t,u)$ with $p=0,t,u=1,r=2,s=-1$, and a projection as a member of $G_2(k)$ with $k = 2$.

\

Families $G_1(k)$ and $G_2(k)$ fall within family $F(p,q,r,s,t,u)$. We present these explicitly as they are a natural way to visualise some products of recombination mediated by a tyrosine recombinase.

\

Note that not all knots and catenanes in family $F(p,q,r,s,t,u)$ are predicted to arise as products of recombination. In particular, $F(p,q,r,s,t,u)$ contains catenanes with up three components. However, it is impossible to yield a three component catenane from recombination on a knot substrate. Figure 11 show the exact topology of the products predicted. Notice that they are all knots or catenanes with up to two components.


\section{Assumptions of our Model}\label{assumptions}
 Given a twist knot substrate and a given recombinase, we now state assumptions about the recombinase-DNA complex. Evidence that these assumptions are biologically reasonable is given in Section 2 of \cite{BFbio}.

\begin{assumptionB} The recombinase complex is a productive synapse, and there is a projection of the crossover sites with 0 or 1 crossing between the sites and no crossings within a single site.
\end{assumptionB}

 Figure \ref{projectionprerec} illustrates projections of $B$ before recombination. 

\begin{assumptionB} The productive synapse does not pierce through a supercoil or a branch point in a nontrivial way and the supercoiled segments are closely juxtaposed. Also, no persistent knots or catenanes are trapped in the branches of the DNA on the outside of the productive synapse.
\end{assumptionB}

 Figure \ref{prodsynp} illustrates examples of recombinase complexes that either are or are not productive synapses.
Figure \ref{ass2} illustrates different examples of DNA molecules that are and that are not allowed according to  assumption 2. Note that we allow hooked junctions (see Figure \ref{crossingconvention}) because these have projections where there is only one crossing between the sites, but no projections with no crossings between the sites. 

\vspace{.3cm}
 \textbf{Assumption 3 for Serine recombinases.} \textit{Serine recombinases perform recombination via the \textbf{subunit exchange mechanism} \cite{bioreview}. This mechanism involves making two simultaneous double-stranded breaks in the sites, rotating two recombinase monomers in opposite sites by $180^{\circ}$ within the productive synapse and resealing the new DNA partners. In each subsequent round of processive recombination, the same set of subunits is exchanged and the sense of rotation remains constant.}

\vspace{.3cm}
 Figure \ref{ass3s} illustrates Assumption 3 for serine recombinases.
It illustrates projections of $B$ at each round of processive recombination mediated by a serine recombinase. Recall that in processive recombination, the term substrate refers specifically to the DNA prior to the first cleavage. 

\vspace{.3cm}
 \textbf{Assumption 3 for Tyrosine recombinases.} \textit{ After recombination mediated by a tyrosine recombinase, there is a projection of the crossover sites which has 0 or 1 crossing.}
\vspace{.3cm}

 Figure \ref{ass3t} illustrates Assumption 3 for tyrosine recombinases.
It illustrates all possible projections of $B$ after recombination mediated by a tyrosine recombinase. For the post-recombinant synapse, note that we allow hooked junctions  because these have projections where there is only one crossing between the sites, but no projections with no crossings between the sites. Note also that tyrosine recombinases can sometimes give the appearance of ``processive recombination'' in the circumstances involving multiple rounds of recombine and reset, without loss of accessory factor binding. In this work we regard this action mediated by tyrosine recombinases as distributive recombination.


\section{Results}\label{results}

 Given the three assumptions in the previous section, we predict that all product knots and catenanes of (non-distributive) site-specific recombination on twist knots with a tyrosine recombinase (Theorem 1) or with a serine recombinase (Theorem 2) fall within the one family of knots and catenanes illustrated in Figure \ref{fams}$(a)$. We also  predict the exact knot and catenane type of possible products of one round of recombination on a twist knot substrate that have MCN one more than the substrate molecule. The technical proofs of these results can be found in \cite{KDmaths}. 

\subsection{Products of non-distributive site-specific recombination belong to three families of knots and catenanes}
\begin{theoremm} \textbf{(Tyrosine recombinases)} Suppose that Assumptions 1, 2 and 3 hold for a particular tyrosine recombinase-DNA complex. Then the only possible products of (non-distributive) recombination on a twist knot $C(2, v)$ are those illustrated in the left half of Figure \ref{theoremssummary}.

\

(That is, if the substrate knot is a twist knot $C(2,v)$ then the only possible products (of a non-distributive reaction) are the unknot, the two node catenane $2_1^2$, $T(2,m)$ for $m=v,v\pm1,v\pm2$,  $C(2,s)$ for $s=v\pm1,v\pm2$, $C(r,v)$ for $r=\{\pm2,\pm3,4\}$, a connected sum $T(2,\pm2)\sharp C(2,v)$, a member of the family $F(p,q,r,s,t,u)$ with $r=2,|t|\leq2$ ,$p=0$, or a member of the family of knot and links $G_1(k)$ or of the family of knots and links $G_2(k)$.)
\end{theoremm}

\begin{theoremm} \textbf{(Serine recombinases)} Suppose that Assumptions 1, 2 and 3 hold for a particular serine recombinase-DNA complex. Then the only possible products of $n$ rounds of processive (non-distributive) recombination on a twist knot $C(2, v)$ are those illustrated in right half of Figure \ref{theoremssummary}.

\

(That is, if the substrate knot is $C(2,v)$ then the only possible products of $n$ rounds of (non-distributive) processive recombination are the $C(r,v)$ for $r=\pm n, \pm n+2$, $C(2,s)$ for $s=v,v\pm n$, $T(2,v\pm n)$, a connected sum $T(2,\pm n)\sharp C(2,v)$ and any member of the family $F(p,q,r,s,t,u)$ with $r=2,t=\pm n$ and $p=0$.) 
\end{theoremm}

 \textbf{Note:} Theorems 1 and 2 distinguish between the chirality of the product DNA molecules, since using our model we can work out the 
\textit{exact conformation of all possible products} of site-specific recombination starting with a particular twist knot substrate and site-specific recombinase. For example, starting with the twist knot substrate $C(2,-1)$ (a $(-)$ trefoil, see Figure \ref{RLTrefoil} for an illustration of a $(-)$-trefoil and $(+)$-trefoil)
then according to our model site-specific recombination mediated by a tyrosine recombinase yields $T(2, -5)$, which is a $(-) 5_1$ (among other products) and can never yield $T(2,+5)$, which is a $(+) 5_1$. 
%
%


\subsection{Characterization of products of distributive recombination}

\begin{theoremm} Any products whose knot or catenane type is not listed in the Theorems 1 and 2 must arise from distributive recombination.
\end{theoremm}

 \textit{Knots and links in $F(p,q,r,s,t,u)$ that cannot arise from recombination mediated  by a serine recombinase, nor} by a tyrosine recombinase. Recall that all products from recombination with a tyrosine recombinase or a serine recombinase belonging to $F(p,q,r,s,t,u)$ can be expressed with $t=\pm n, p=0, r=2$. Thus, any knots that cannot be expressed in this form cannot arise as products of recombination with either a serine recombinase or a tyrosine recombinase. $8_{18}$ and $10_{141}$ are examples of such knots.



\

\textit{Knots that can arise as products of recombination mediated by a serine recombinase, but not by a tyrosine recombinase}. In contrast with Theorem 1, any knot or link in the family $F(p,q,r,s,t,u)$ with $|t|>2, r=2, p=0$, can occur as a consequence of Theorem 2. The knot $F(0,-1,2,-1,3,-1)=8_{13}$ mentioned above is an example of this; this knot is a possible product of recombination with a serine recombinase, but not with a tyrosine recombinase.

\subsection{Products whose MCN is one more than the substrate}

Often recombination increases the minimal crossing number of a knotted
or catenated substrate by one (e.g., \cite{11}).  In this case, we  can further restrict the knot and catenane type of the possible products of recombination.

\begin{theoremm}\label{OneMoreMCNThm} Suppose that Assumptions 1, 2, and 3 hold for a particular recombinase-DNA complex with substrate $J=C(2,v)$ and suppose site-specific recombination increases the minimal crossing number by one. 
Let $L$ be the product of a single recombination event. 
Then for $v>0$, $L$ can be any of the knots and catenanes illustrated in the left half of Figure \ref{onemoreMCN} and for $v<0$, $L$ can be any of the knots and catenanes illustrated in the right half of Figure \ref{onemoreMCN}. These are the only possibilities for $L$.

\

(That is, if $v>0$, $L$ is one of the following: $C(2,v+1)$, $C(2,-v)$, $C(-2,v)$, $C(-2,-(1+v))$, $C(3,v)$, $T(2,\pm(2+v))$, $F(0,q,2,s,2,u)$ where $q+s=v$, $F(0,\pm1,2,s,0,u)$ where $q+s=v$ and $s\neq0$ or $F(0,0,2,s,2,u)$ where $q+s=v$.

\

If $v<0$ $L$ is one of the following: $C(2,2+|v|)$, $C(2,-(1+|v|))$, $C(-2,1+|v|)$, $C(-2,-(2+|v|))$, $C(3,v)$, $C(-4,v)$, $T(2,\pm(3+|v|))$ or $F(0,\pm1,2,s,0,u)$ for $q+s=v$.)
\end{theoremm}



\section{Discussion and Applications} \label{applications}
Our model predicts products of processive and distributive recombination in a number of ways: We outline three algorithms that help us:  predict all possible products of (non-distributive) processive site-specific recombination on twist knot substrates mediated by a serine recombinase, determine the sequence of products of processive recombination given a substrate and a list of experimentally characterized products, and determine all the products of (non-distributive) recombination of a twist knot substrate, mediated by a tyrosine recombinase. We illustrate how to use these algorithms in Applications 1, 2 and 3. In Application 3 we we employ Algorithm 3 to analyse previously uncharacterised products of distributive recombination mediated by a tyrosine recombinase.   In Application 4 we discuss how our model can reduce the number of possibilities of products in situations where they have MCN one more than the substrate. In Application 5 we explain how our model can be used to distinguish between products of processive and distributive recombination.

\subsection{Algorithm 1: Determining all products of recombination mediated by a serine recombinase}\label{algorithm3}
\begin{enumerate}
\item Choose a twist knot substrate. Determine if the substrate is of the form $C(-2,v)$ or $C(2,v)$. If necessary, employ a deformation of the substrate molecule by flipping the hook to make the substrate molecule the twist knot $C(2,v)$ (instead of $C(-2,v-1)$). Determine $v$.
\item  The right half of Figure \ref{theoremssummary}  illustrates all the possible conformations of products of site-specific recombination on the twist knot $C(2,v)$ mediated by a serine recombinase. For each, replace $v$  with the value of $v$ (that is, $v$ should be replaced for the value of $v$ of the substrate. For instance, if the substrate is the trefoil knot $C(2,-1)$ then the number of crossings in the row of crossings $v$ in these figures should be replaced by one negative crossing) and recall that $q+s=v$ so consider each combination of $q$ and $s$ such that $q+s=v$.
\item For a substrate that does not have mismatched sites: to obtain all possible products of the $k^{\mbox{th}}$ round of processive recombination, for each each image of the right half of Figure \ref{theoremssummary}, replace $n$ by $k$  either positive, negative crossings if the handedness of recombination of the enzyme is known, or both positive and negative crossings if the handedness is not known. 
\item For a substrate that has mismatched sites: to obtain all products of the $k^{\mbox{th}}$ round of processive recombination, replace $n$ by $2k$ crossings.
\item Exclude any predicted products that are not possible due to the relative orientation and alignment of the specific sites or any other special properties of the particular system in question. (Depending on the relative orientation (direct or inverted repeat) of the sites catenanes may or may not be possible products of the reaction. Also, the alignment (parallel or antiparallel) of the sites, may produce mismatched crossover sites. In this case, an even number of site exchanges are necessary in order for the recombinase to be able to reseal the DNA sites, leaving the parental genomic sequence intact but introducing two or more crossings.)
\end{enumerate}

\textit{\textbf{Application 1.} Predicting all products of recombination on a twist knot substrate mediated by a serine recombinase.} Suppose we choose the substrate $C(2,-2)$ with wild-type sites. Then $v=-2$, so replace $v=-2$ in each image of the right-hand side of Figure \ref{theoremssummary}. Suppose also that recombination proceeds through one round of strand exchange during each round of processive recombination and that we do not know the direction of rotation of one half of the recombinase complex relative to the other, during strand exchange. We want to know all possible products of the first three rounds of processive recombination. 

\

The products of the first round of recombination (obtained by replacing $n$ with one (positive and negative) crossing) are: $C(+1,-2)$, $C(-1,-2)$, $C(+1+2,-2)$, $C(-1+2,-2)$, $C(2,-2)$, $C(2,-2+1)$, $C(2,-2-1)$, $T(2,-2+1)$, $T(2,-2-1)$, $T(2,+1)\sharp C(2,-2)$, $T(2,-1)\sharp C(2,-2)$, $F(0,0,2,-2,+1,u)$, $F(0,0,2,-2,-1,u)$, $F(0,-1,2,-1,+1,u)$, $F(0,-1,2,-1,-1,u)$, $F(0,-2,2,0,+1,u)$, $F(0,-2,2,0,+1,u)$, $F(0,q,2,-2,+1,u)$, $F(0,q,2,-2,-1,u)$.

\

The products of the second round of recombination (obtained by replacing $n$ with two (positive and negative) crossings) are: $C(+2,-2)$, $C(-2,-2)$, $C(+2+2,-2)$, $C(-2+2,-2)$, $C(2,-2+2)$, $C(2,-2-2)$, $T(2,-2+2)$, $T(2,-2-2)$, $T(2,+2)\sharp C(2,-2)$, $T(2,-2)\sharp C(2,-2)$, $F(0,0,2,-2,+2,u)$, $F(0,0,2,-2,-2,u)$, $F(0,-1,2,-1,+2,u)$, $F(0,-1,2,-1,-2,u)$, $F(0,-2,2,0,+2,u)$, $F(0,-2,2,0,+2,u)$, $F(0,q,2,-2,+2,u)$, $F(0,q,2,-2,-2,u)$.

\

The products of the third round of recombination (obtained by replacing $n$ with three (positive and negative) crossings) are: $C(+3,-2)$, $C(-3,-2)$, $C(+3+2,-2)$, $C(-3+2,-2)$, $C(2,-2+3)$, $C(2,-2-3)$, $T(2,-2+3)$, $T(2,-2-3)$, $T(2,+3)\sharp C(2,-2)$, $T(2,-3)\sharp C(2,-2)$, $F(0,0,2,-2,+3,u)$, $F(0,0,2,-2,-3,u)$, $F(0,-1,2,-1,+3,u)$, $F(0,-1,2,-1,-3,u)$, $F(0,-2,2,0,+3,u)$, $F(0,-2,2,0,+3,u)$, $F(0,q,2,-2,+3,u)$, $F(0,q,2,-2,-3,u)$.

\

These correspond to:  (knots) $0_1, (-)3_1, 4_1, 5_2, (-)6_1, 7_2$, (catenanes) $(+)2^2_1, (-)4^2_1$, (composite knots and catenanes) $(\pm3_1)\sharp 4_1,(\pm)2^2_1\sharp4_1$ and any products belonging to the infinite families $F(0,0,2,-2,\pm n,u)$, $F(0,-1,2,-1,\pm n,u)$, $F(0,-2,2,0,\pm ,u)$, $F(0,q,2,-2,\pm n,u)$ for $u$ any positive integer. Two important observations: the products not in the infinite families are very tightly prescribed and the handedness of all the chiral products (except $2^2_1$) is negative.

\subsection{Algorithm 2: Determining the sequence of products of processive recombination (given a specific substrate and experimentally characterized products)}\label{algorithm1}

When the products of processive recombination are known, but not the ordering, the algorithm below can determine the sequence of products.

\begin{enumerate}
\item Determine if the substrate is of the form $C(-2,v)$ or $C(2,v)$.
If necessary, employ a deformation of the substrate molecule by flipping the hook to make the substrate molecule the twist knot $C(2,v)$ (instead of $C(-2,v-1)$). Determine $v$.
\item Figure \ref{serineproducts} lists all the possible conformations of the synaptic complex and the possible products that particular conformations yields. Use the characterised products from the experiment to determine which image in Figure \ref{serineproducts} is the the conformation of the synaptic complex. Perform the rest of the algorithm only for that conformation.
\item For Figures $\ref{serineproducts}(b)-\ref{serineproducts}(h)$ replace $v$ by its value (that is, $v$ should be replaced for the value of $v$ of the substrate). In Figure $\ref{serineproducts}(a)$ recall that the number of crossings $q$ and $s$ should add up to the number of crossings $v$. So consider each combination of $q$ and $s$ such that $q+s=v$.
\item  Figure \ref{serineproducts} also gives the possible pre-recombinant  forms of $B$ for each conformation of the synaptic complex, and Figure \ref{ass3s} gives the post-recombinant forms of $B$ at each stage of processive recombination. Use the characterised products  and the images in Figure \ref{serineproducts} to determine the pre-recombinant form of $B$ and for that form, use Figure \ref{ass3s}, to work out which path yields those products.
\item In the synaptic complex form chosen, replace $B$ with each of its post-recombinant forms at each stage of processive recombination to obtain a sequence of products.
\item Exclude any predicted products that are not possible due to specific properties of the particular system in question. (See the step $(5)$ of the algorithm in Section \ref{algorithm3}.)
\end{enumerate}

This algorithm gives the order of products of processive recombination and any products arising from the experiment that are not in the list of predicted products are assumed to arise from distributive recombination.

\


\textit{\textbf{Application 2.} Determining the possible sequences of products of processive recombination.}

\textbf{\textit{Example.}} Suppose that for the twist knot substrate $C(-2,3)$, experimental conditions minimize distributive recombination and analysis of the products reveal unknots, (unknown) torus knots and (unknown) clasp knots $C(r,s)$.

\

We can determine the order of products of recombination by using Figures  \ref{ass3s} and \ref{serineproducts} as follows:
\begin{itemize}
\item We perform a deformation of the substrate molecule by flipping the hook to make the substrate molecule the twist knot $C(2,4)$ (instead of $C(-2,3)$).
\item From Figure \ref{serineproducts} we see that the conformation of the synaptic complex must be that illustrated in Figure \ref{serineproducts}(e). This is the only conformation of the synaptic complex that yields unknots, torus knots (and catenanes) and clasp knots. 
\item Also from Figure \ref{serineproducts}, the pre-recombinant conformation of $B$ must be form $B1$.
\item From Figure \ref{ass3s}, we see that path 1 is the only path that gives an unknot as a product (of the second round of recombination). 
\end{itemize}

Thus, in Figure \ref{serineproducts}(e) replace the $B$ by the post-recombinant form after each round of processive recombination. We deduce that the sequence of products is: $C(2,4)$ (substrate) $\rightarrow$ $T(2,-1+4)$ $\rightarrow$ $C(0,4)$ (unknot) $\rightarrow$ $T(2,1+4)$ $\rightarrow$ $C(-2,4)$ $\rightarrow$ $C(-3,4)$. Any products of further rounds of processive recombination are clasp knots $C(r,4)$ with increasing minimal crossing number. 

\

\textbf{\textit{Example (Substrate with mismatched sites).}} 
In cases where there is a mismatch in the crossover sites, two subunit exchanges are necessary in order for the recombinase to be able to reseal the DNA sites. (That is, processive recombination performs two $180^{\circ}$ rotations of one half of the productive synapse relative to the other before ligating the sites.)

\

 Assume that for the twist knot substrate $C(-2,2)$ experimental conditions minimize distributive recombination and that processive recombination encounters mismatched sites after one round of exchange, so it proceeds through two exchanges of the crossover sites per round of recombination. Suppose that the products of multiple rounds of processive recombination are twist knots and connected sums of a torus knot and a twist knot. Then we employ a a similar method to that explained in the example above:

\begin{itemize}
\item We perform a deformation of the substrate molecule by flipping the hook to make the substrate molecule the twist knot $C(2,3)$ (instead of $C(-2,2)$).
\item Figure \ref{serineproducts}(c) gives the conformation of the synaptic complex because this is the only conformation that gives connected sums of torus knots or catenanes and twist knots.
\item $B$ is either $B3$ or $B4$ because: from \ref{serineproducts}(c), $B=B1, B3$ or $B4$, but all paths of $B1$ give only the twist knot $C(2,3)$.
\item  The path must be path 1 for both $B3$ and $B4$ because path 2 does not give any twist knots and paths 3 and 4 give only the twist knot $C(-2,v)$.
\end{itemize}

So, using $B=B3$, path 1, we obtain the following sequence of products: 
$C(2,3) \mbox{ (substrate) } \rightarrow C(2,3) \rightarrow T(2,-3)\sharp C(2,3) \rightarrow T(2,-5)\sharp C(2,3)$.
Moreover, any products of further rounds of recombination are connected sums of the form $T(2,m)\sharp C(2,3)$ for $m$ an odd negative integer, with increasing minimal crossing number. 

\

Using $B=B4$, path 1, we obtain the following sequence of products: 
$C(2,3) \mbox{ (substrate) } \rightarrow C(2,3) \rightarrow T(2,3)\sharp C(2,3) \rightarrow T(2,5)\sharp C(2,3)$
 Moreover, any products of further rounds of recombination are connected sums of the form $T(2,m)\sharp C(2,3)$ for $m$ an odd positive integer, with increasing minimal crossing number.

\subsection{Algorithm 3: Determining all products of (non-distributive) recombination mediated by a tyrosine recombinase}\label{algorithm2}
\begin{enumerate}
\item Use the characterised products from the experiment to determine if the substrate is a twist knot of the form $C(2,v)$ or $C(-2,v)$. If necessary, employ a deformation of the substrate molecule by flipping the hook to make the substrate molecule the twist knot $C(2,v)$ (instead of $C(-2,v-1)$). Determine $v$.

\item  The left half of Figure \ref{theoremssummary},  illustrates all the possible conformations of a product of site-specific recombination on the twist knot $C(2,v)$ mediated by a tyrosine recombinase. For products other than $G_1(k)$ or $G_2(k)$, replace $v$ (and/or $q$ and $s$, recall $q+s=v$, so consider all combinations) with the value of $v$ (that is, $v$ should be replaced for the value of $v$ of the substrate).
\item For products of the form $G_1(k)$ or $G_2(k)$:
\begin{itemize}
\item If $v$ is a negative odd number, the product is a knot belonging to family $G_1(k)$ with $k = |v|$.
\item If $v$ is a positive odd number, the product is a knot belonging to family $G_2(k)$ with $k = |v|-1$.
\item If $v$ is a negative even number, the product is a link belonging to family $G_1(k)$ with $k = |v|$.
\item If $v$ is a positive even number, the product is a link belonging to family $G_2(k)$ with $k = |v| - 1$.
\end{itemize}

\item Exclude any predicted products that are not possible due to specific properties of the particular system in question. (See the step $(5)$ of the algorithm in Section \ref{algorithm3}.)
\end{enumerate}

\

%

\noindent For both serine and tyrosine site-specific recombinases,  any product molecules that arise from the experiment that are not in the list of knots and catenanes  predicted by the algorithms are assumed to have arise from distributive recombination reactions.

\

\textbf{\textit{Application 3.}} \textit{Narrowing possible knot and catenane type for previously uncharacterised experimental data: Products of distributive recombination of a tyrosine recombinase.} We now consider products of multiple rounds of distributive recombination mediated by a tyrosine recombinase. In \cite{10} Crisona \textit{et al} performed experiments using the Flp recombinase of the yeast $2 \mu$m plasmid on unknotted substrates. They studied Flp inversion reactions by carrying out the experiments on plasmids containing two inverted FRT sites. Note that FRT sites in inverted repeats cannot yield catenanes as products of one round of recombination with a Flp recombinase. Flp can catalyse multiple rounds of distributive recombination, so it forms both even and odd-noded knots. In their paper, they were only interested in the odd-noded knots as they are products of the first round of distributive recombination. They found that trefoil knots $C(2,-1)=3_1$ were among these products and did not identify product of further rounds of distributive reactions.

\

An interesting question is: What are the possible products of further rounds of distributive recombination? This can be answered using  Theorem 1 and Figure \ref{theoremssummary} (left-hand side). We assume that the product of the first round of distributive recombination (and thus our substrate) is the trefoil knot $C(2,-1)$. In Figure \ref{theoremssummary} (left-hand side), we set $v=-1$ in each image (since the substrate is $C(2,-1)$). 
From this we can see that the possible products of a second round of distributive recombination by Flp recombinase on the unknot are 
$0_1, 3_1,4_1, 5_1, 5_2, 0^2_1, 2^2_1, 4^2_1, 2^2_1\sharp3_1$ and any non-catenane products belonging to the family $F(p,q,2,s,t,u)$ with with $p=0, q+s=-1, |t|\leq 2$, as illustrated in Figure \ref{theoremssummary} (left-hand side). (Recall that the catenanes predicted are not biologically possible.) (Note that here we did not take into account handedness, but it can be revealed by writing the products in $C(r,s)$ and $T(2,m)$ notation as in the left-hand side of Figure \ref{theoremssummary}.) Recall that for these images, the vertical row of $u$ crossings can be any number of crossings. This means that topologically, there are infinitely many possibilities for these product knots. However, biologically, due to physical and other constrains of the DNA molecule,  conformations  with a small value of $u$ would probably be the most abundant.

\subsection{Other applications of our model}\label{apps}

\

\textit{\textbf{Application 4:} Products of recombination reactions that increase the MCN by 1.} Very commonly, site-specific recombination adds one crossing to the substrate, resulting in an increase by one of the MCN of the substrate. For example Bath \textit{et al} used the catenanes $T(2,6)$ and $T(2,8)$ as substrates for XerCD recombination yielding product knots with MCN equal to 7 and 9 \cite{73}. They did not characterise these products beyond their MCN (although they did make strong biological predictions as to their exact topology). Buck and Flapan \cite{BFbio} significantly reduced the possibilities for each of these products, Darcy \cite{Darcy} used the tangle model to reduce the number of mathematical solutions to the tangle equations involving the 4-noded catenane $4^2_1$ (product of one round of recombination on an unknot substrate) and a 7-noded knot (products of one round of recombination on the torus catenane $6^2_1$) and Vazquez \textit{et-al} \cite{vazquez-et-al}  designed a three-dimensional model for Xer recombination. 

\

As DNA twist knots are common recombination substrates, considering a similar scenario to the Xer example above is relevant. To do this we  apply Theorem \ref{OneMoreMCNThm}. Figure \ref{onemoreMCN} summarizes this  theorem. Suppose the twist knots $C(2,5)$ and $C(2,7)$ (MCN 6 and 8 respectively) are used as substrates for a site-specific recombination reaction with Xer recombinase, where experimental conditions minimize distributive recombination  and products are knots and catenanes with minimal crossing number 7 and 9. (Note that Xer has been shown to recombine on both directly and inversely repeated sites, thus knots or catenanes are expected depending on site orientation.) In this case the minimal crossing number is not sufficient to determine the knot type, since there are 7 knots and 8 two-component catenanes with MCN=7 and 49 knots and 61 two-catenanes with MCN=9. However, we can use  Theorem \ref{OneMoreMCNThm} (and Figure \ref{onemoreMCN}) to significantly reduce the number of possibilities for these products. By replacing $v$ for $5$ crossings in each illustration of Figure \ref{onemoreMCN}, it follows  that the possible seven-crossing products are $7_1$, $7_2$, $7_3$, $7_6$, $7^2_2$, $7_3^2$, or $3_1\sharp4_1$. Similarly, by replacing $v$ for $7$ crossings in each illustration of Figure \ref{onemoreMCN}, it follows  that the possible nine-crossing products are $9_1$, $9_2$, $9_3$, $9_8$, $9_{11}$, $9^2_1$, $9^2_{10}$, $6_1\sharp 3_1$, or $4_1\sharp 5_2$. We have reduced from 15 choices for 7-noded knots to just 7 and from 110 possibilities for 9-noded knots and catenanes to just 9 possibilities.  Thus, Theorem \ref{OneMoreMCNThm} can help to significantly reduce the knot and catenane type of products of site-specific recombination that add one crossing to the substrate.

\

\textit{\textbf{Application 5:} Processive vs Distributive recombination.} In some cases, processive recombination does not preclude distributive rounds of recombination, and both occur in a recombination reaction. Our model can be helpful in distinguishing between products of distributive recombination and products of processive recombination. 

\

\textbf{\textit{Example.}} Suppose that a trefoil knot $C(-2,1)$, is used as a substrate for a reaction with a serine recombinase and that electron microscopy and gel electrophoresis reveal the figure of eight knot $C(-2,2)$ as the primary product and $T(2,2)$$\sharp$$C(-2,1)$,  $T(2,2)$ $\sharp$ $C(-2,2)$ and a three-component catenane as secondary products. It follows from  Theorem 2 that recombination proceeds from the trefoil knot to $C(-2,2)$, product of the first round of processive recombination. The original $C(-2,1)$  and the product $C(-2,2)$ are then substrates yielding the composite catenanes $T(2,2)$$\sharp$$C(-2,1)$ and $T(2,2)$$\sharp$$C(-2,2)$, products of the first round of distributive recombination. The product knots and these composite catenanes are then used as substrates to yield the three-component catenanes, products of the second round of distributive recombination (however, this composite is not one of the substrates that we consider). Overall, this would be akin to the serine recombinase performing multiple rounds of processive and distributive recombination.

\section{Conclusions and Directions for Further Research.}
 We have developed a model to predict and characterise the topology and chirality of DNA knots and catenanes that can arise as a result of a
site-specific recombinase acting on a twist knot substrate. Our model is based
on three biological assumptions about site-specific recombination. Our model predicts that all knotted or
catenated products of such enzyme actions are in one of the three families of Figure \ref{fams}, as described in Theorems 1 and 2.

\

 In \cite{KDmaths} we have also shown that the total number
of knots and catenanes in our product families  with MCN$=n$ grows linearly with $n^5$, whereas the total number of all knots and catenanes increases exponentially with the MCN \cite{ErntsSumnersGrowthOfNumberOfPrimeKnots}. Hence, the calculation $n^5/e^n$ gives the proportion of all knots and catenanes which are putative recombination products and as $n$ increases, $n^5/e^n$ decreases exponentially rapidly to zero. Knowing the MCN of a product and knowing that
the product is in one of our families allow us
to significantly narrow the possibilities for its knot
or catenane type. The model described herein thus
provides an important step in characterizing DNA
knots and catenanes, which arise as products of site-specific
recombination.

\

We have shown how our model can be helpful in determining the sequences of products of processive recombination on twist knot substrates, and how it can help distinguish between products of processive and distributive recombination. The algorithms presented allow the interested reader to apply our results to a specific site-specific recombination system.

\

We plan to expand this project in three main ways. First, we are further testing the model with experimental data from our experimental collaborators. Second, we are developing a computer program based on the model presented on this paper and in \cite{KDmaths}. This will allow the automatic computation of products of site-specific recombination on any twist knot substrate.
Finally, although we have assumed that the productive synapse has only two crossover sites and that any accessory sites are sequestered from the synaptic complex, electron micrographs of recombinase complexes such as those of Gin and Hin \cite{5,6,7,8,mutranspososome} reveal three three DNA duplexes looping out of the enzyme complex. This suggests that our model could be developed by making biologically reasonable assumptions of a synaptic complex with three crossover sites (see for example \cite{3tangle}) and predicting the putative products that could arise.

\section{Acknowledgments} 
 We wish to thank Erica Flapan,  Mauro Mauricio, Julian Gibbons, Kai Ishihara, Ken Baker for insightful discussions, and the referees for their valuable suggestions.  DB is supported in part by EPSRC Grants EP/H0313671, EP/G0395851 and EP/J1075308, and thanks the LMS for their Scheme 2 Grant.  KV is supported by EP/G0395851.



\end{twocolumn}

\begin{onecolumn}
\clearpage
\section*{Figures}

\begin{figure}[h]
\begin{center}
\includegraphics[width=16cm]{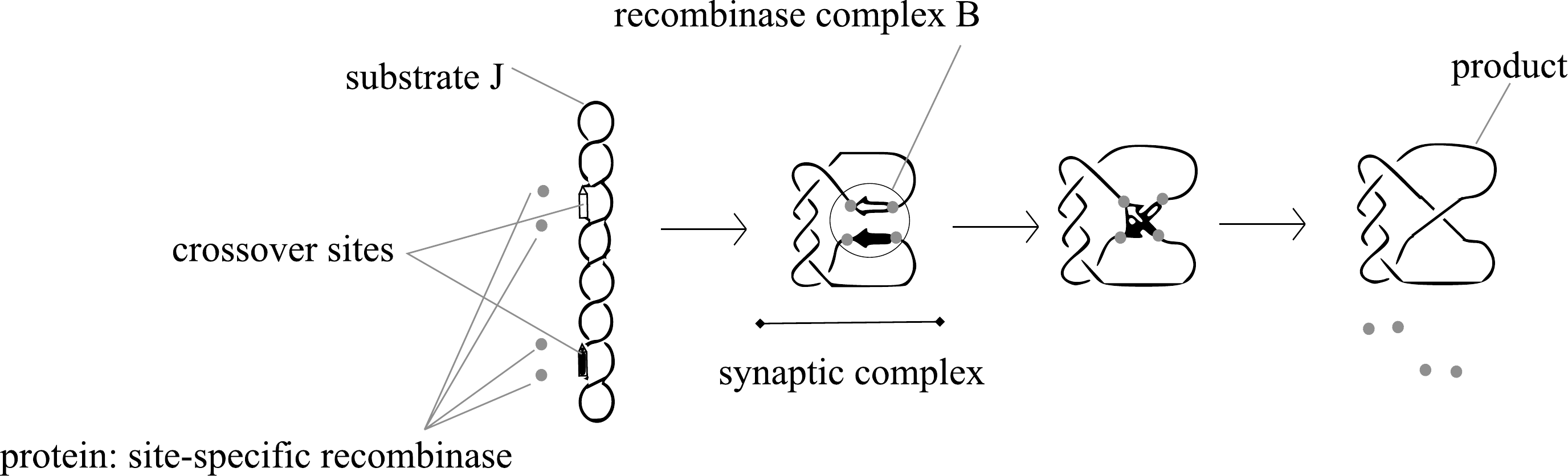}
\caption{Illustration of a small site-specific recombination reaction. The black line represents the central axis of the DNA molecule. \textit{Left:} Four dimer site-specific recombinases. The substrate molecule: an unknotted, duplex covalently closed-circular and plectonemically supercoiled DNA molecule with the crossover sites (empty and filled arrows).  \textit{Left middle:} Juxtaposition of the crossover sites, creating the synaptic complex. \textit{Right middle:} Recombination of the crossover sites. \textit{Right:} Knotted product molecule and detached proteins. } 
\label{SSRsummary}
\end{center}
\end{figure}

\begin{figure}[h]
\begin{center}
\includegraphics[width=.8\textwidth]{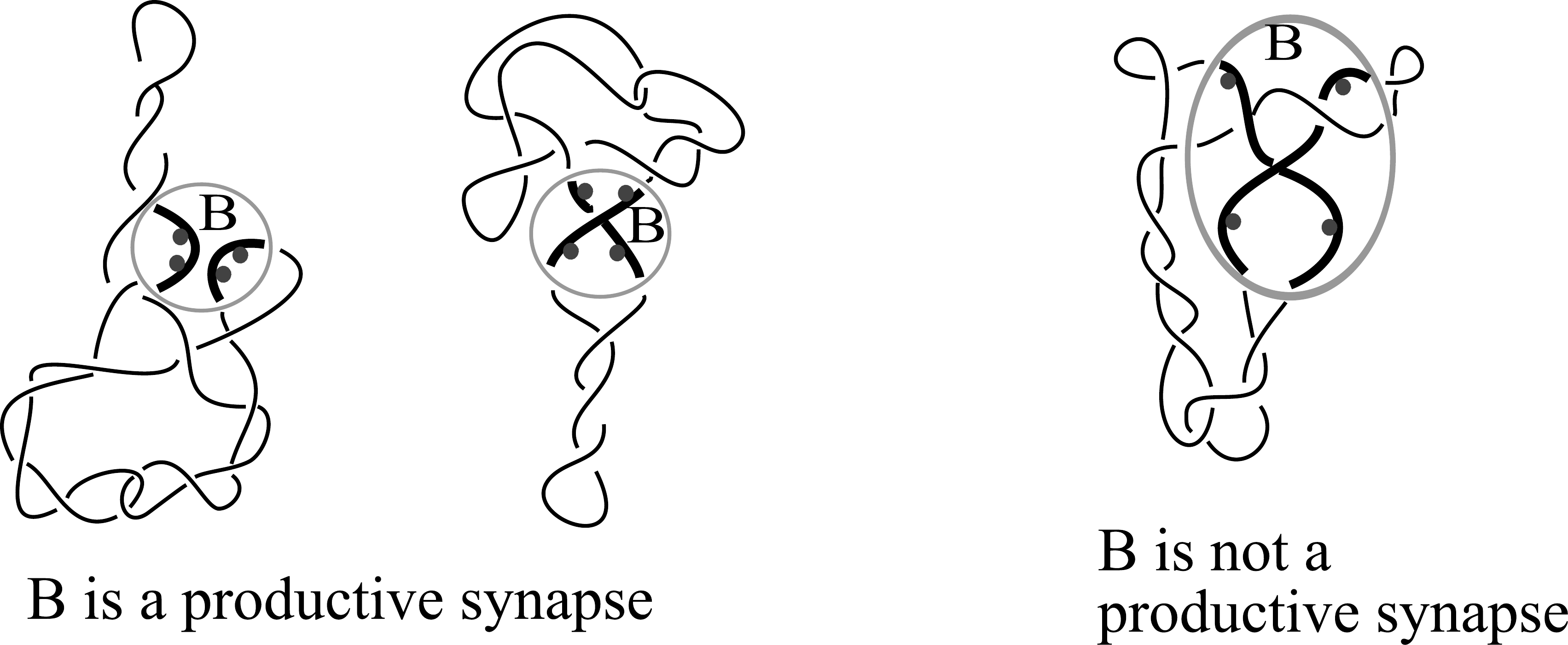}
\caption{Productive synapse. We assume that the synaptic complex is a productive synapse.  $B$ (light grey circle) denotes the smallest  region containing the four bound recombinase molecules (small grey discs) and the two crossover sites (highlighted in black). \textit{Left and middle:} The synaptic complex is a productive synapse. \textit{Right:} The synaptic complex is not a productive synapse. In this case we cannot draw $B$ such that only the two crossover sites are inside it without also including the third (horizontal, non-highlighted) strand.}
\label{prodsynp}
\end{center}
\end{figure}

\begin{figure}[h]
  \centering
\subfloat[][]{\label{crossingconvention}\includegraphics[width=.4\textwidth]{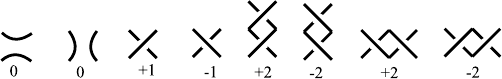}}
  \hspace{0.2cm}
 \subfloat[][]{\label{linkknot}\includegraphics[width=.3\textwidth]{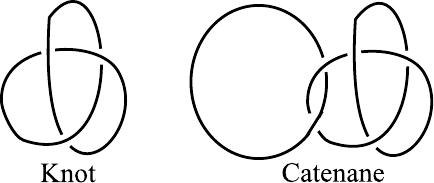}}
  \hspace{0.2cm}
 \subfloat[][]{\label{TorusKnott}\includegraphics[width=.1\textwidth]{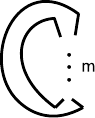}}
  \hspace{0.2cm}
 \subfloat[][]{\label{substrate}\includegraphics[width=.1\textwidth]{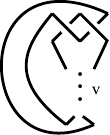}}
  \hspace{0.2cm}
 \subfloat[][]{\label{claspknot}\includegraphics[width=.1\textwidth]{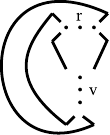}}
    \hspace{0.2cm}
       \subfloat[][]{\label{compositeknot}\includegraphics[width=.17\textwidth]{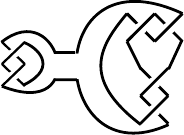}}
  \hspace{0.2cm}
   \subfloat[][]{\label{RLTrefoil}\includegraphics[width=.25\textwidth]{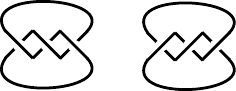}}
\caption{Background terminology. 
\textit{(A)} Crossing sign convention used in this paper: $0$ horizontal and vertical; $+1,-1$; $+2$ and $-2$ vertical; $+2$ and $-2$ horizontal, also called hooked junctions.    \textit{(B)} A example of a knot and a catenane. \textit{(C)} Generalised structures of torus knots ($m$ is odd) and catenanes ($m$ even) denoted $T(2,m)$.\textit{(D)} The substrate we consider, the twist knot $C(2, v)$. \textit{(E)} A clasp knot $C(r, s)$. \textit{(F)} An example of a composite catenane. This particular example, denoted $T(2,2)\sharp C(-2,2)$ consists of the catenane $T(2,2)=2^2_1$ and the twist knot $C(-2,2)=4_1$.  \textit{(G)} The $(+)$trefoil and $(-)$trefoil, respectively from left to right.
}
\label{defs}
\end{figure}

\begin{figure}[h]
\begin{center}
\includegraphics[width=8cm]{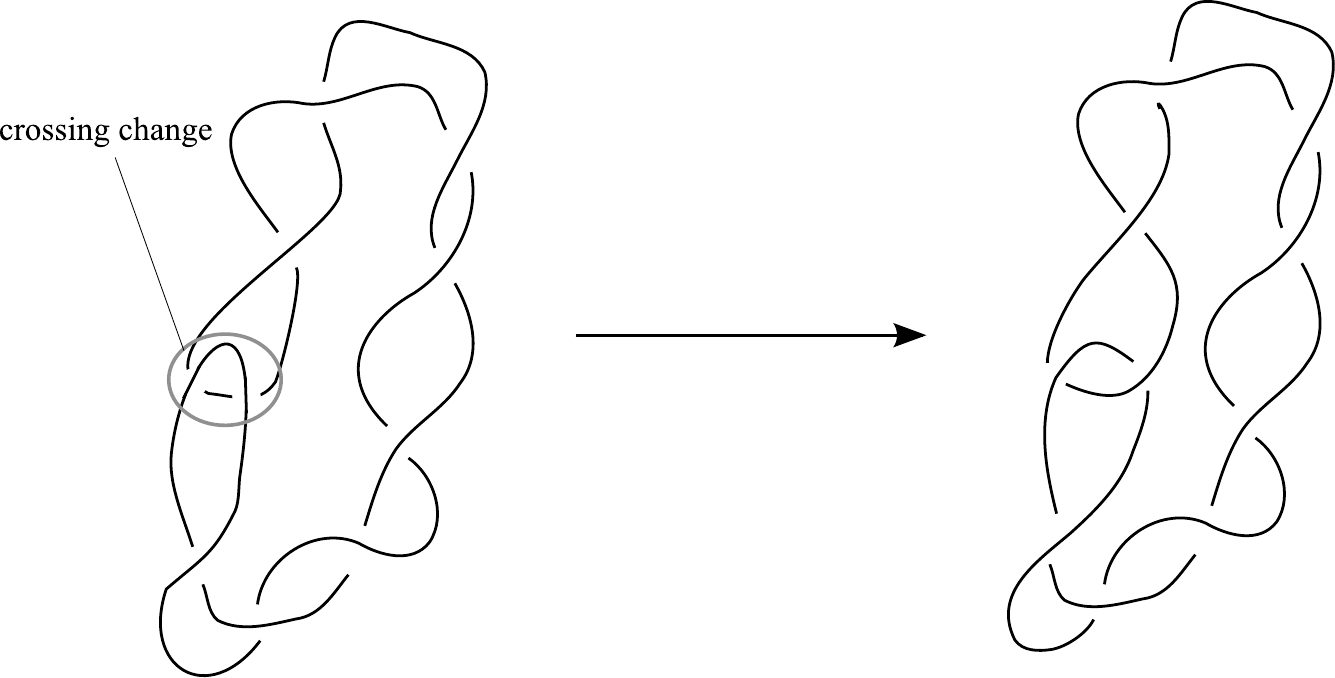}
\caption{Twist knots are ubiquitous DNA knots. DNA \textit{in vivo} and \textit{in vitro} is plectonemically supercoiled so an unknot can be transformed to a twist knot by a single crossing change.}
\label{supercoiledtotwist}
\end{center}
\end{figure}

\begin{figure}[h]
\begin{center}
\includegraphics[width=8cm]{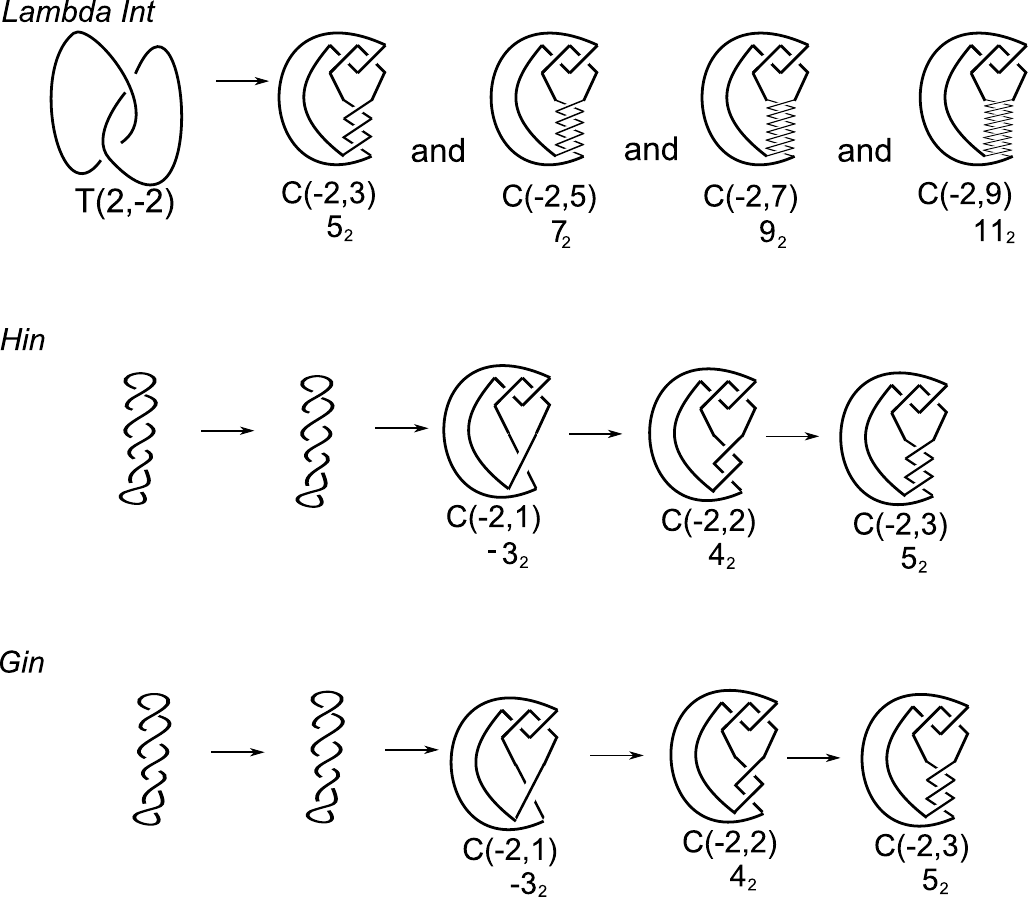}
\caption{Site-specific recombination mediated by both serine recombinases and tyrosine recombinases often yields twist knot products. }
\label{Page3Exampless}
\end{center}
\end{figure}

\begin{figure}[h]
  \centering
    \subfloat[][]{\label{famsF}\includegraphics[width=.3\textwidth]{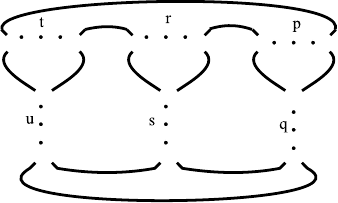}}
      \vspace{0.5cm}
  \subfloat[][]{\label{famsG1}\includegraphics[width=.2\textwidth]{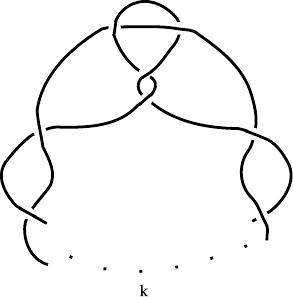}}
  \vspace{0.5cm}
  \subfloat[][]{\label{famsG2}\includegraphics[width=.2\textwidth]{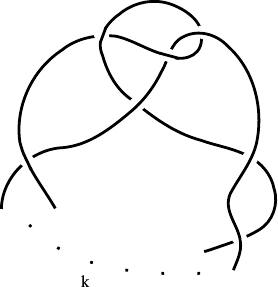}}
\caption{The family of knots and catenanes $F(p,q,r,s,t,u)$, containing the subfamilies $G_1(k)$ and $G_2(k)$  (from left to right respectively). Given the three assumptions in Section \ref{assumptions}, we predict that all product knots and catenanes of (non-distributive) site-specific recombination on twist knots with a tyrosine recombinase or a serine recombinase fall within family $F(p,q,r,s,t,u)$ of knots and catenanes. In \ref{famsF}, the letters $p,q,r,s,t$ and $u$ denote the number of crossings in that particular row of crossings.  In \ref{famsG1} and \ref{famsG2}, $k$ describes the number of crossings between the two DNA duplexes. Note that depending on the value of $k$, a member of $G_1(k)$ of $G_2(k)$ is either a knot or catenane. $G_1(k)$ and $G_2(k)$ are important subfamilies of $F(p,q,r,s,t,u)$ which arise as products of recombination mediated by a tyrosine recombinase. }
\label{fams}
\end{figure}

\begin{figure}[h]
\begin{center}
\includegraphics[width=.4\textwidth]{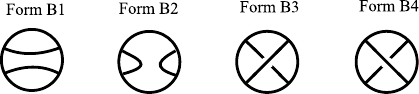}
\caption{Assumption 1:  Projections of the pre-recombinant $B$. Assumption 1 states that there is a projection of the pre-recombinant recombinase complex with at most one crossing.  }
\label{projectionprerec}
\end{center}
\end{figure}



\begin{figure}[h]
\begin{center}
\includegraphics[width=12cm]{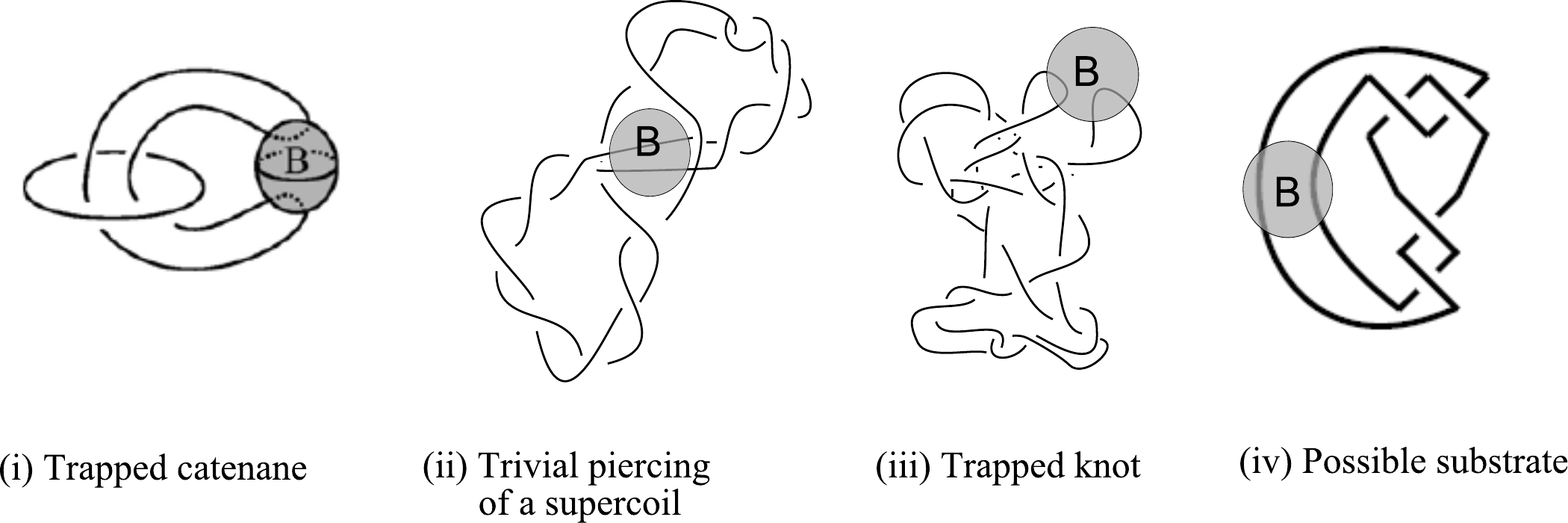} 
\caption{Different scenarios for Assumption 2. \textit{(i)} A catenane is trapped in the DNA branches outside of $B$. \textit{(ii)} The productive synapse pierces a supercoil in a non-trivial way. \textit{(iii)} A knot is trapped in the DNA branches outside of $B$. \textit{(iv)} An unknotted substrate with the synaptic complex already formed. Scenarios \textit{(i)} and \textit{(iii)} are not allowed by the assumption, \textit{(ii)} and \textit{(iv) are.} }
\label{ass2}
\end{center}
\end{figure}

\begin{figure}[h]
\begin{center}
\includegraphics[width=18.3cm]{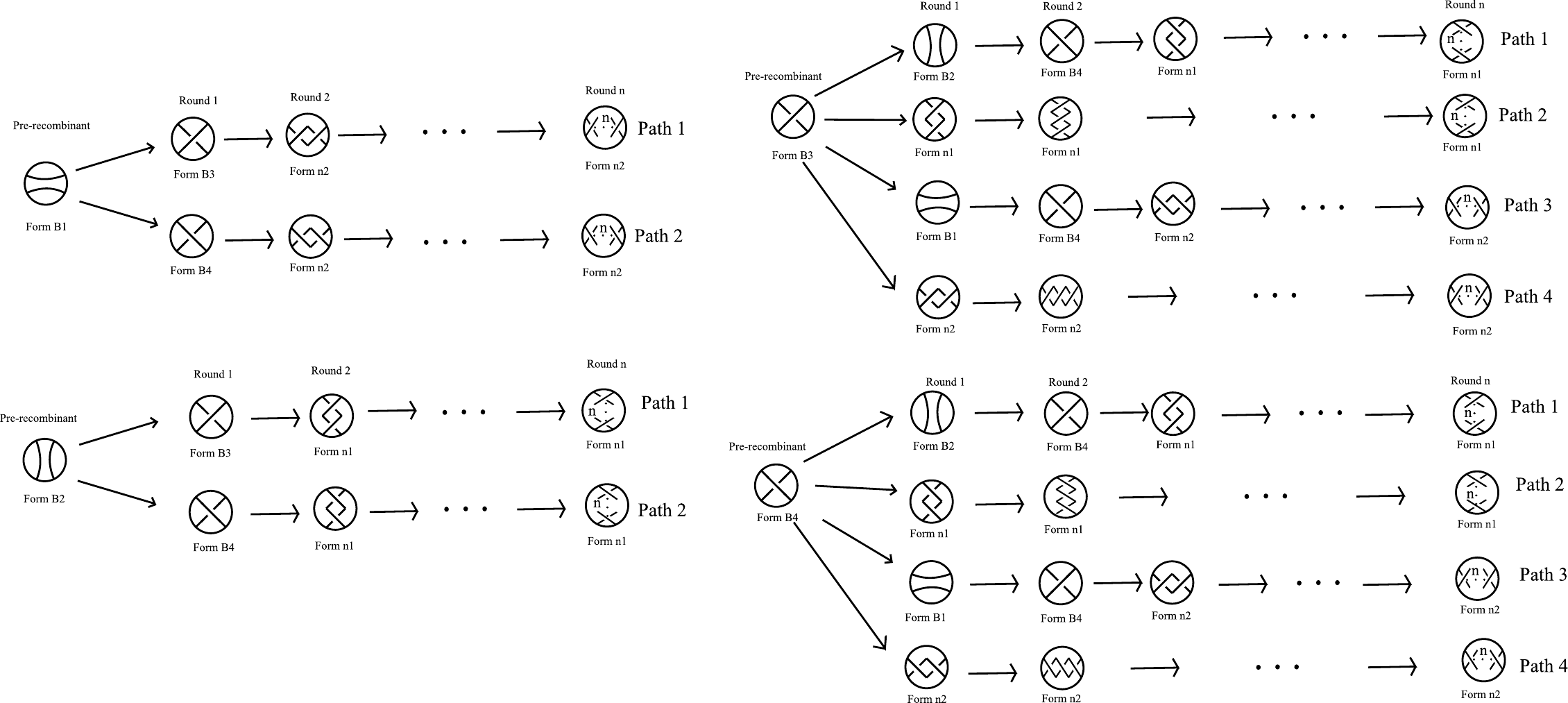} 
\caption{Assumption 3 for Serine recombinases. Starting with a projection of pre-recombinant $B$ with zero or one crossings, we illustrate projections of the post-recombinant conformations of $B$ at each round of processive recombination. Processive recombination can result in a row of $n$ vertical crossings, which we denote by $n1$ or in a row of $n$ horizontal crossings which we denote $n2$.}
\label{ass3s}
\end{center}
\end{figure}

\begin{figure}[h]
\begin{center}
\includegraphics[width=14cm]{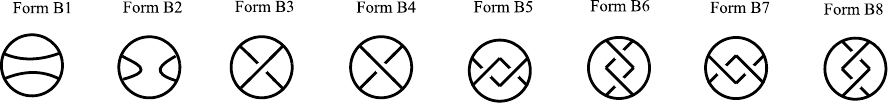} 
\caption{Assumption 3 for Tyrosine recombinases. All possible projections of $B$ after recombination mediated by a tyrosine recombinase.}
\label{ass3t}
\end{center}
\end{figure}

\begin{figure}[h]
\begin{center}
\includegraphics[width=18cm]{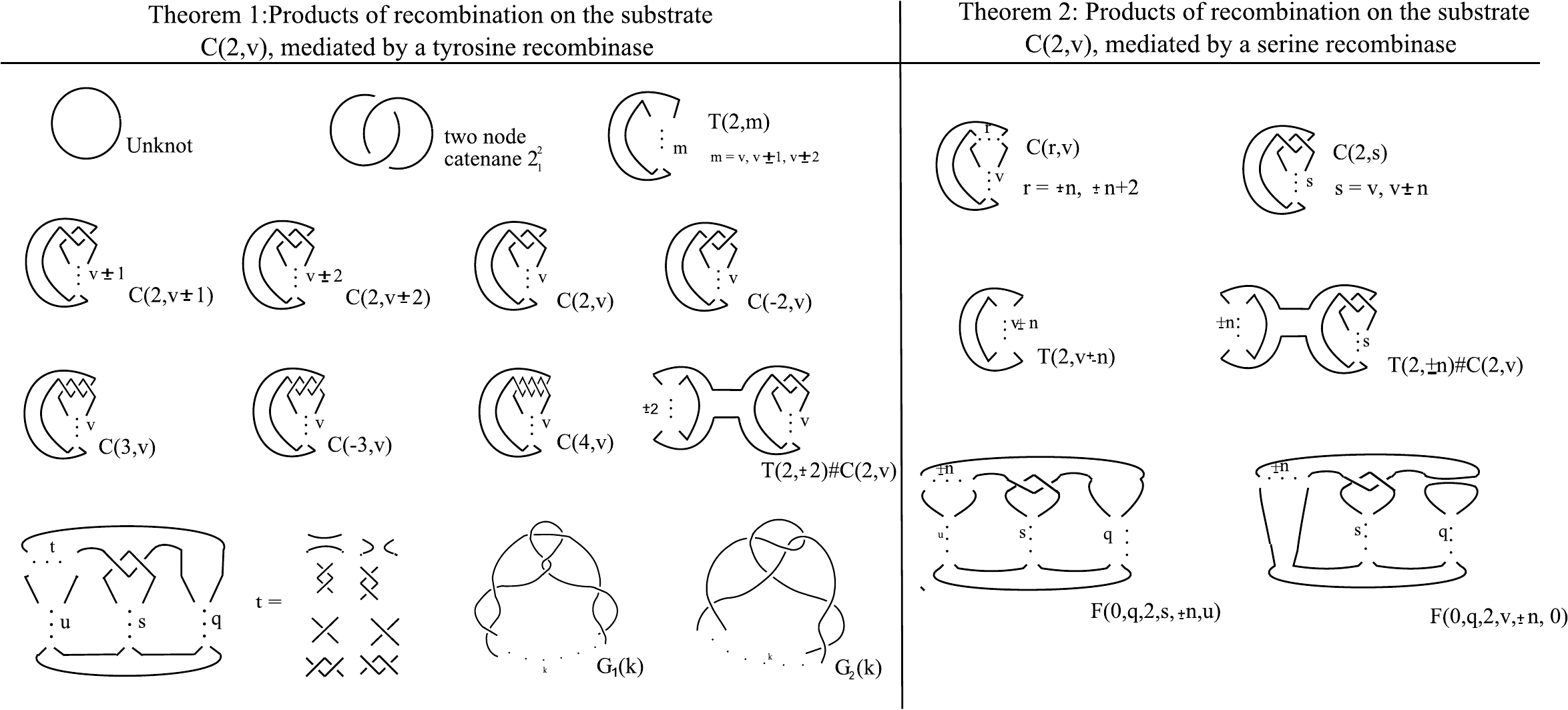} 
\caption{\textit{Left:} Summary of Theorem 1. These are all the possible products of a reaction mediated by a tyrosine recombinase on a twist knot substrate $C(2,v)$, predicted by the model. \textit{Right:}  Summary of  Theorem 2. Similarly, these are all the possible products of a reaction mediated by a serine recombinase on a twist knot substrate $C(2,v)$, predicted by the model. For all, $q+s=v$.These products are listed in the same order as in the Theorems (left to right, top to bottom). Note that all products predicted are either knots or catenanes with up to two components. On the right half of the image, $n$ is an integer.}
\label{theoremssummary}
\end{center}
\end{figure}

\begin{figure}[h]
\begin{center}
\includegraphics[width=18.5cm]{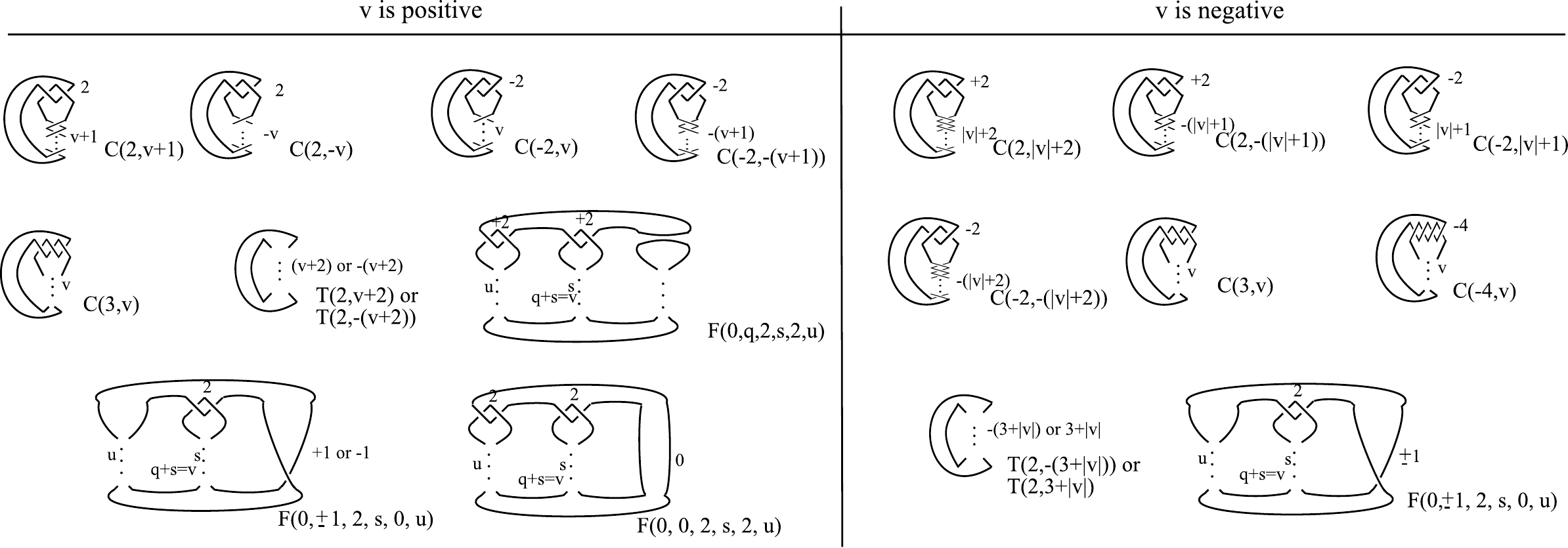} 
\caption{Summary of Theorem \ref{OneMoreMCNThm}. Products of a recombination reaction with a twist knot substrate that have MCN one more than that of the substrate.  If the substrate is a twist knot $C(2,v)$ with MCN$(C(2,v))=m$ and the product has MCN equal to $m+1$ then the knots and catenanes illustrated here are the only possible such products. Depending on whether $v$ is positive or negative (see illustration \ref{defs} for the convention on crossings) then we obtain different possible products. For all, $q+s=v$. These products are listed in the same order as in the Theorems (left to right, top to bottom).}
\label{onemoreMCN}
\end{center}
\end{figure}

\begin{figure}[h]
\begin{center}
\includegraphics[width=13cm]{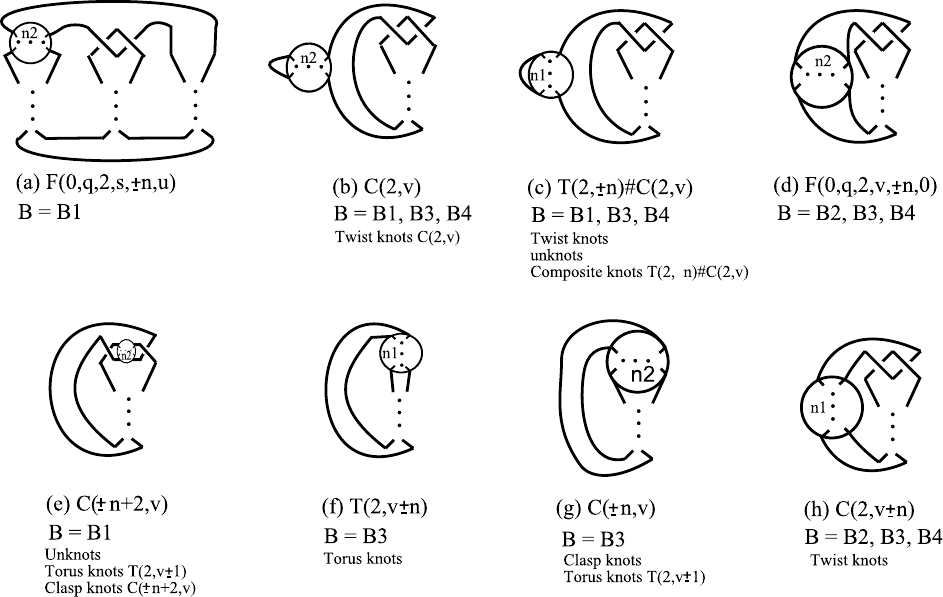} 
\caption{Summary of Theorem 2: After $n$ rounds of processive recombination with a serine recombinase on a twist knot substrate, the product DNA molecule must have one of the forms illustrated here.  The images inside the circles denote  $B$ after $n$ rounds of processive recombination. There are two possible conformations, a vertical row of $n$ crossings, $n1$ and a horizontal row of $n$ crossings, $n2$. Under each conformation we list some of the possible products that can arise from that particular conformation after $n$ rounds of processive recombination.}
\label{serineproducts}
\end{center}
\end{figure}



\end{onecolumn}
\end{document}